# Resonant high-order harmonic generation by indium ions

ALEXANDER I. MAGUNOV[1,2,3] AND VASILY V. STRELKOV[1,2,4]

[1]*Prokhorov General Physics Institute of the Russian Academy of Sciences, 38 Vavilova Street, Moscow 119991, Russia*
[2]*Institute of Applied Physics of the Russian Academy of Sciences, 46 Ulyanov Street, Nizhny Novgorod 603950, Russia*
[3]*State Scientific Center "National Research Institute for Physical–Technical and Radiotechnical Measurements", Mendeleevo, Moscow region, 141570, Russia*
[4]*Moscow Institute of Physics and Technology (State University), Dolgoprudny, Moscow Region 141700, Russia*
*strelkov.v@gmail.com

**Abstract:** High-order harmonic generation in multiphoton ionization regime is studied theoretically to examine the resonant effects of the generating particle. We solve the time-dependent Schrödinger equation for $In^+$ in the laser field by expanding the solution in terms of unperturbed eigenstates obtained in the Hartree-Fock approximation. The intensity enhancement of the high harmonic is demonstrated under 7-photon resonance with the energy of the transition from the ground state to the $4d^9 5p$ $^1P_1$ autoionizing state. Attosecond pulse generation using harmonics of the fifth and higher orders is studied. We find a train of almost spectral limited pulses with duration of 190 as.

## 1. Introduction

High-order harmonic generation in intense laser field is actively studied both experimentally and theoretically. This is a perspective approach for production of coherent collimated XUV of femtosecond or attosecond duration. Current state of these studies is presented in reviews [1-4].

Majority of experimental studies use noble gas jets as generating media. However, recently HHG using other targets is actively studied as well. Such targets are atomic clusters, nanoparticles, tin films, and others. In particular, many experiments investigate HHG in laser plumes [5-9]. In these studies first the heating laser pulse produces plasma plume at a surface of a solid target. Then the intense femtosecond laser pulse generates XUV in the plasma. An advantage of this approach in comparison with gaseous HHG is much wider choice of the generating particle type. In particular, the harmonic generation is observed using ions having strong dipole transitions at frequencies resonant with the harmonic ones. The resonant character of the process leads to highly-efficient HHG. The impressive results are achieved using singly-charged metallic ions. The resonant enhancement of the efficiency (i.e. the ratio of the efficiency of the resonant harmonic to efficiency of the neighboring ones) of the 13[th] harmonic of Ti:Sapphire fundamental (800 nm) in indium plasma plume is up to two orders of magnitude, and the enhancement for the 7[th] harmonic generation from 400 nm field (the second harmonic of Ti:Sapphire laser) in gallium plasma is almost three orders of magnitude [7].

The most pronounced resonant effects were observed when the harmonic frequency is resonant with the frequency of a transition from the ground to an autoionizing state (AIS, i.e. the quasi-bound state with the excitation energy exceeding the ionization threshold) involving excitation of two electrons of the outer shell, or one electron from the inner shell [10,11]. For the ions under consideration an AIS lifetime is of the order of femtoseconds. The resonant HHG involving resonance with AIS has two advantages. First, it leads to highly-efficient generation of short-wavelength radiation (the photon energy exceeds the ion ionization energy); HHG is one of the very few sources providing such coherent radiation. Second, for some ions the transition from AIS to the ground state has high oscillator strength leading to high efficiency of the emission.

For the generation conditions corresponding to the tunneling ionization (the Keldysh parameter [12] $\gamma < 1$) rather good level of understanding and theoretical description of the resonant HHG is achieved [13-19]. In particular, in [15] the so-called four-step model of the resonant HHG was suggested. The first two steps in this model are the same as ones in the well-known three-step model of the non-resonant HHG [20] (these steps are the electron detachment from the ion and the electron quasi-free motion in the laser field). At the next step the electron is captured by the parent ion so that the whole system finds itself in the AIS. The final step is the transition to the ground state with the XUV photon emission. Note that all the steps are coherent with the laser field, so this process leads to the coherent XUV emission. The theory [17] based on the four-step model describes not only the resonant enhancement of the harmonic generation efficiency, but also the harmonic phase behavior near the resonance.

For the conditions corresponding to the multiphoton ionization regime, or to the intermediate regime ($\gamma \geq 1$) the understanding level of the HHG process is lower. Early papers ([21,22] and many others) considered relatively low laser intensities and relatively low nonlinearity orders. The resonant HHG in gallium plasma for the multiphoton regime was studied in [9] using numerical solution of the time-dependent Schrödinger equation (TDSE) for a model ion in single-active electron approximation. The pronounced enhancement of the resonant harmonic generation was found, although lower than one observed experimentally. Further numerical studies were done for a model two-electron system in [23]. Note that a drawback of such approach is lack of actual atomic/ionic level structure reproduction (energies of the levels, oscillator strengths, autoionization and photoionization widths). This partly concerns models based on Slater determinants with term averaged energies [18,24,25].

In this paper we study theoretically the HHG spectra for a singly-charged indium ion. The study is based on the expansion of TDSE solution for ionic wave-function in terms of a set of stationary bound states. The time-dependent expansion coefficients (transition amplitudes) are determined numerically from a system of ordinary differential equations. The spectroscopic characteristics of the ion are calculated using the Hartree–Fock self-consistent field approximation [26] and being corrected if necessary in comparison with available data in the NIST collection [27]. In our calculation we pay particular attention to the role of multiphoton resonances in the enhancement of certain harmonic intensities.

The atomic units ($e = m_e = \hbar = 1$) are used throughout the paper unless otherwise stated.

## 2. Method

In the semiclassical approach the photon emission spectral probability by an atom in the intense laser field with a frequency $\omega$ and unit polarization vector **e** is defined by:

$$P(\omega) = \frac{4}{3\omega c^3} \left| \mathcal{F}_\omega[\ddot{\mathbf{D}}_{av}(t)] \cdot \mathbf{e} \right|^2. \tag{1}$$

Here $\mathcal{F}_\omega[\ddot{\mathbf{D}}_{av}(t)]$ is the Fourier transform of the second derivative of the quantum mechanical average of the dipole moment vector found as $\mathbf{D}_{av}(t) = \langle \Psi(t)|\mathbf{D}|\Psi(t)\rangle$, and $\Psi(t)$ is a wave function found solving the TDSE:

$$i\,\partial_t \Psi(t) = \left[\hat{H} - \mathbf{D} \cdot \mathbf{F}(t)\right]\Psi(t), \tag{2}$$

where $\hat{H}$ is the Hamiltonian of an unperturbed atom, $\mathbf{D} = -\sum_j \mathbf{r}_j$, $\mathbf{r}_j$ is the $j$-electron coordinate, $\mathbf{F}(t)$ is the laser field strength. The latter is written as $\mathbf{F}(t) = F_p f(t) \,\mathrm{Re}[\mathbf{e}_L \exp(-i\omega_L t)]$ where $F_p$ is the peak value, $f(t)$ is the normalized envelope function, $\omega_L$ is the carrier frequency, and $\mathbf{e}_L$ is the unit polarization vector (real for the linear polarization). We suppose the long wavelength approximation, thereby neglecting the spatial dependence of the field strength.

In our study the solution of (2) is expanded in a set of eigenstates $\varphi_n$ of $\hat{H}$ (generally including continuum):

$$\Psi(t) = \sum_n a_n(t)\exp(-iE_n t)\varphi_n,  \quad (3)$$

where $E_n$ is the eigenenergy. The initial conditions for the expansion coefficient $a_n(t)$ corresponding to the initial state of an atom $\varphi_i$ read $a_n(t \to -\infty) \to \delta_{ni}$.

Combination of (2) and (3) leads to the set of equations for the expansion coefficients:

$$\dot a_n(t) = i \sum_{n'} \exp(i\omega_{nn'}t)\mathbf{D}_{nn'} \cdot \mathbf{F}(t) a_{n'}(t),  \quad (4)$$

where $\mathbf{D}_{nn'} = \langle \varphi_n|\mathbf{D}|\varphi_{n'}\rangle$ is the dipole matrix element and $\omega_{nn'} = E_n - E_{n'}$ is the transition frequency.

The average dipole moment is expressed in terms of the expansion coefficients as follows:

$$\mathbf{D}_{av}(t) = 2\mathrm{Re}\sum_{nn'} \exp(i\omega_{nn'}t)\mathbf{D}_{nn'} a_n^*(t) a_{n'}(t),  \quad (5)$$

where $n$ and $n'$ stand for even and odd atomic states, respectively.

Practically series (3) of a finite size should be used. We consider discrete states and AISs shown in the scheme in Fig. 1.

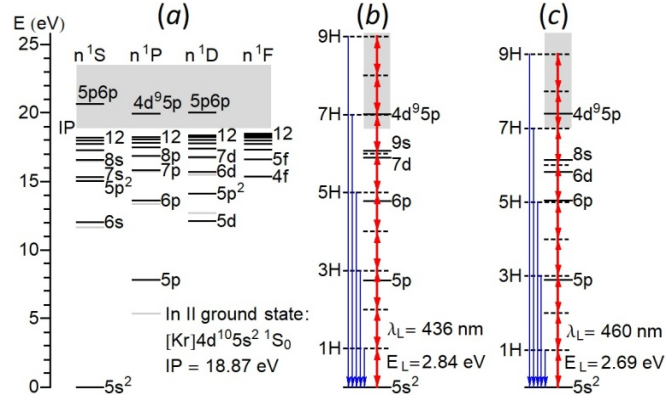

Fig. 1. (a) Scheme of the singlet levels of the In$^+$ ion which are accounted for the HHG calculation. Gray area denotes continuum states. Triplet levels of the nearest configuration are indicated by grey lines. Multiphoton transitions and nearest intermediate states at laser wavelength 436 nm with a close resonance of the 7-th harmonic and the AIS 4d$^9$5p $^1$P$_1$ (b), a nonresonant condition at the wavelength 460 nm (c). Red arrows are the laser induced transitions; blue arrows are the harmonic emission. Dashed lines are the virtual states.

Only singlet states with the angular momentum projection $M = 0$ on the direction of laser field linear polarization are taken into account. The triplet states with much lower transition strength from the ground state may be important only under resonant conditions, which can be controlled. The omitted higher excited state can by strongly affected by environment in plasma target media that needs a special consideration beyond the scope of the present paper.

Besides the selected states with corresponding dipole bound-bound transitions [28], we take into account the continuum states as intermediate ones in the second order interaction using, which leads to single-photon laser ionization and autoionization widths (corresponding energy shifts are neglected). Note that the multiphoton ionization of lower states by the laser field is partly included via their coupling to higher excited levels.

Finally, the equation set for the state coefficients at the linearly polarized laser field reads:

$$\dot a_n(t) = -\tfrac{1}{2}\Gamma_n(t) a_n(t) - if(t)\cos\omega t \sum_{n'} \exp(i\omega_{nn'}t)\Omega_{nn'} a_{n'}(t),  \quad (6)$$

where $\Omega_{nn'} = \langle \varphi_n|Z|\varphi_{n'}\rangle F_p$ is the Rabi frequency, $\Gamma_n(t) = A_n + \Gamma_n^i(t) + \Gamma_n^a$ is the total relaxation width of the level $n$, $\Gamma_n^a = 2\pi|W_{n,Ec}|^2$ is the autoionization width due to configuration interaction $\widehat{W}$ between $n$ and adjacent continuum state $Ec$, $\Gamma_n^i(t) = \sigma_n^i(\omega_L)I_p[f(t)]^2/\omega_L$ is the laser ionization width of the state $n$ with the cross section $\sigma_n^i(\omega_L)$ under the peak laser intensity $I_p = F_p^2$. The radiation width $A_n$ is also included for the excited states using the procedure described, e.g., in [29]. Usually, $\Gamma_n^a \gg A_n + \Gamma_n^i(t)$ holds.

The decay in equations (6) provides $\mathbf{D}_{av}(t \to \infty) \to 0$, $\dot{\mathbf{D}}_{av}(t \to \infty) \to 0$. In this case $\mathcal{F}_\omega[\ddot{\mathbf{D}}_{av}(t)] = -\omega^2 \mathcal{F}_\omega[\mathbf{D}_{av}(t)]$. Taking this into account equation (1) is written as:

$$P(\omega) = \frac{4\omega^3}{3c^3} |\mathcal{F}_\omega[\mathbf{D}_{av}(t)] \cdot \mathbf{e}|^2. \quad (7)$$

Numerical solution of (6) was performed with the spectroscopic constants for $In^+$ obtained using the Hartree-Fock wave functions [26] and compared with available data from the NIST collection [27] for renormalization to more accurate values.

The envelope function was taken in the form $f(t) = \sin^2(\pi t/n_c T_c)$ at the interval from 0 to $n_c T_c$, where $T_c$ is the optical cycle duration. The laser field is linearly polarized. The dipole moment is found numerically within the pulse duration; further dynamics of the dipole moment is calculated analytically using the states' amplitudes achieved at the end of the laser pulse. Correspondingly, the spectrum consists of the emission within the laser pulse and the post-pulse emission. The latter is denoted below as spontaneous emission of XFID (X-ray free induction decay).

Note that our approach based on system (6) solution can be understood as a generalization of the two-level model for HHG simulation [30,31].

## 3. Results

Examination of an actual multiphoton transition scheme may be very helpful for better understanding of the most important processes during HHG in real atomic system with its specific spectrum. For the $In^+$ ion at the considered laser wavelength values the corresponding schemes are shown in Fig. 1(b,c).

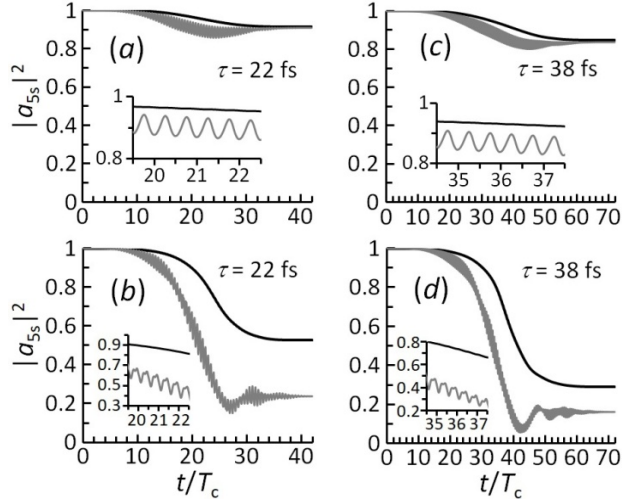

Fig. 2. Population of the ground state $5s^2$ ("noisy" dark gray) and total population of the discrete states (smooth black) as functions of time for different laser pulse durations and peak intensities. (a) Pulse duration (FWHM) $\tau = 22$ fs and peak intensity $I_p = 7.5 \times 10^{13}$ W cm$^{-2}$; (b) $\tau = 22$ fs and $I_p = 1.5 \times 10^{14}$ W cm$^{-2}$; (c) $\tau = 38$ fs and $I_p = 7.5 \times 10^{13}$ W cm$^{-2}$; (d) $\tau = 38$ fs and $I_p = 1.5 \times 10^{14}$ W cm$^{-2}$.

First, we consider the laser wavelength 436 nm (Fig. 1(b)) for which the energy of 7 photons is equal to the transition energy between the ground state and the AIS $4d^95p$ $^1P_1$. Fig. 2 shows the temporal evolution of the ground (initial) level population fraction $|a_{5s}(t)|^2$ and the total population of the discrete states during the pulse for different laser pulse durations and intensities. The inset shows variation in a narrow time range. The Keldysh parameter for the ground state at used wavelength and intensity is more than 2, so the field parameters correspond to the multiphoton ionization regime, rather than to the tunneling one.

In Fig. 2 the black curve presents the total population of discrete states thus the continuum states population is its residue to 1. The excited bound states population is the difference between black and gray curves. In the considered resonance case enhanced population of highly excited levels at the middle of the pulse is confirmed by calculations, especially, at higher intensity shown in Fig. 2(b,d). Population fraction of the continuum states grows with pulse duration and intensity due to direct ionization of highly excited states with ionization threshold below the laser photon energy. At lower intensity (Fig. 2(a,c)) an electron from mostly excited states 5s$n$p returns gradually to the ground state via a kind of multi-photon adiabatic following process.

Since the ground state is most populated among discrete levels even at higher intensity and pulse duration values, the average dipole moment is fed mainly by the odd 5s$n$p states. Fig. 3 shows their population for $n$ = 5-7 and for the AIS 4d$^9$5p $^1$P$_1$ at $I_p$ = 1.5×10$^{14}$ W cm$^{-2}$. The higher $n$ populations are lower at least an order of magnitude.

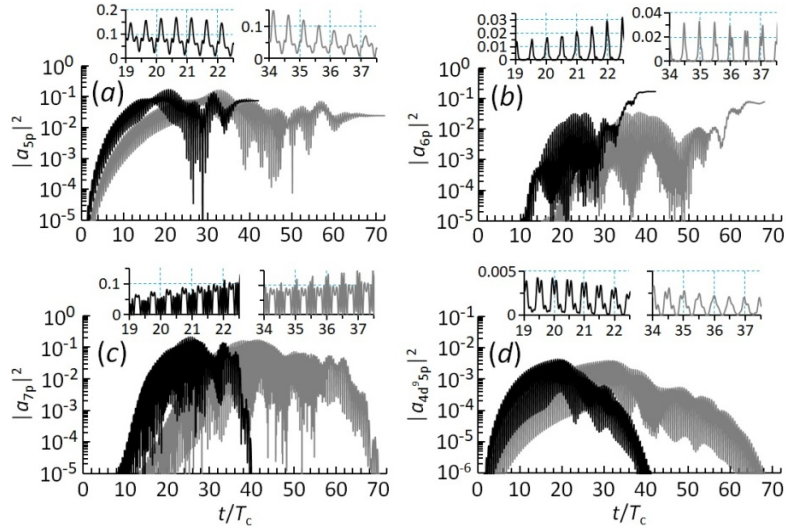

Fig. 3. Population fraction of the 5s$n$p states during the laser pulse with linear polarization at the peak intensity $I_p$ = 1.5×10$^{14}$ W cm$^{-2}$ and the pulse duration values $\tau$ = 22 fs (black) and $\tau$ = 38 fs (dark gray). (a) $n$ = 5 state; (b) $n$ = 6, (c) $n$ = 7, and (d) is the AIS 4d$^9$5p.

Level population behavior is similar for both values of pulse duration. Those for the 5s5p and 5s6p levels stabilize at the end of the pulse nearly on their maximal values for the later level, while the 5s7p level and AIS population in Fig. 3(c,d) follow the pulse envelope. The insets on the top of each plot show a narrow range near the pulse center.

Fig. 4 shows variation of the resonant harmonic 7H line intensity and its neighbors with the laser pulse duration and peak intensity. Narrow red lines illustrate contribution of spontaneous emission (XFID) after the end of a laser pulse according to the residual upper state population (a part of the Fourier integral from $n_cT_c$ to infinity).

Note that the AIS 4d$^9$5p $^1$P$_1$ level population in Fig. 3(d) is small in all cases. It leads to low spontaneous emission from the AIS (red lines at 7H frequency). Only the 6p→5s and 7p→5s (at $\tau$ = 22 fs in Fig. 4(a,b)) transitions lines are observed. The intensity of 7H exceeds those of the neighboring harmonics that manifests the enhancement effect.

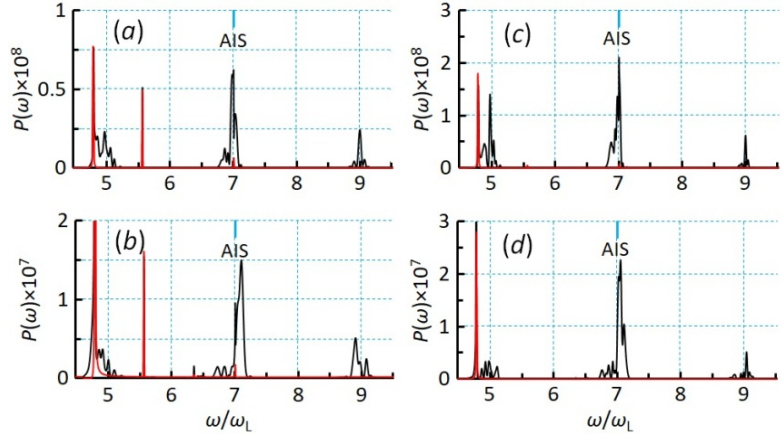

Fig. 4. Harmonic spectra (7) of In$^+$ calculated for the laser wavelength of 436 nm and different laser pulse FWHM durations $\tau$ and peak intensities $I_p$. (a) $\tau$ = 22 fs and $I_p$ = 7.5×10$^{13}$ W cm$^{-2}$; (b) $\tau$ = 22 fs and $I_p$ = 1.5×10$^{14}$ W cm$^{-2}$; (c) $\tau$ = 38 fs and $I_p$ = 7.5×10$^{13}$ W cm$^{-2}$; (d) $\tau$ = 38 fs and $I_p$ = 1.5×10$^{14}$ W cm$^{-2}$. The energy position of the AIS 4d$^9$5p $^1$P$_1$ is indicated by blue line on the top of the panel. Red lines correspond to spontaneous line emission.

To check a resonance effect in HHG we consider the laser wavelength 460 nm with a shift of 7H from the excitation energy of the AIS shown in Fig. 1(c). The results for the harmonic intensities at the same laser pulse parameters are shown in Fig. 5. In this case the 6p→5s transition line with a narrow red peak is close to the 5H on its blue wing. However, this narrow line is unlikely to be observed because of dissipation effects in decaying plasma after the end of the pulse and an instrumental spectral resolution. It also concerns higher-$n$ transitions at $I_p$ = 1.5×10$^{14}$ W cm$^{-2}$ shown to the left of 7H in Fig. 5(b,d).

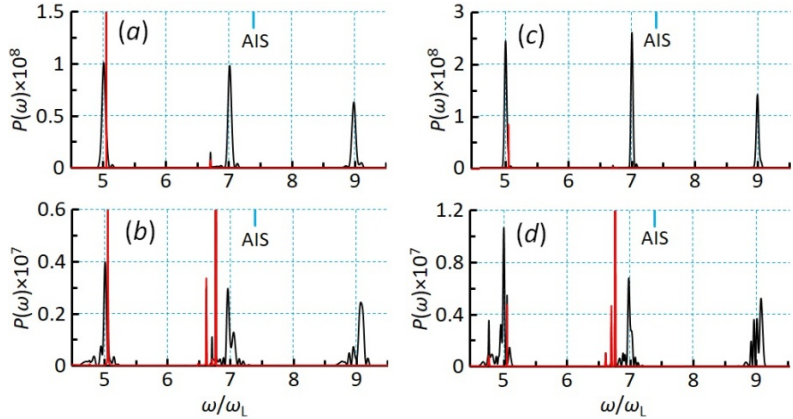

Fig. 5. The same as in Fig. 4, but for the laser wavelength 460 nm. The AIS 4d$^9$5p $^1$P$_1$ (blue line on the top) is shifted from the 7H.

Fig. 6 shows high-$n$ 5s$n$p state populations during the laser pulse for the wavelength 460 nm at the peak intensity 1.5×10$^{14}$ W cm$^{-2}$ and the pulse duration 22 fs and 38 fs. The insets on the top of each plot shows a narrow range behavior near the pulse center. The 5s9p level population at $\tau$ = 22 fs follows the pulse envelope decreasing to a low value at the end of the pulse. The residual population of $n$ = 10-12 states at the end of the pulse produces narrow peaks below 7H in Fig. 5(b,d).

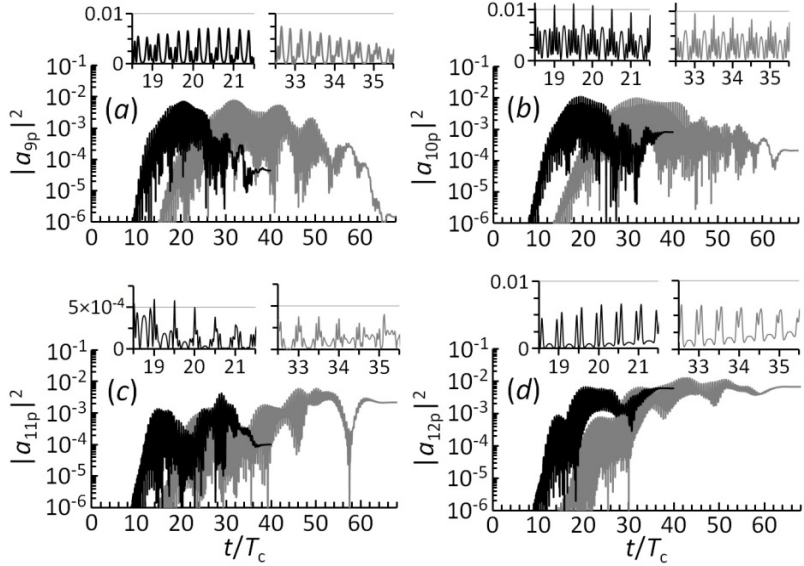

Fig. 6. Population fractions of the high-$n$ 5$sn$p states during the laser pulse at the peak intensity $I_p = 1.5\times10^{14}$ W cm$^{-2}$ and the pulse duration values $\tau = 22$ fs (black) and $\tau = 38$ fs (dark gray).
(a) $n = 9$ state, (b) $n = 10$, (c) $n = 11$, and (d) is $n = 12$.

The population of the ground state and the total population of discrete states during the pulse with different durations are shown in Fig. 7 at the laser wavelength 460 nm and the peak intensity value $1.5\times10^{14}$ W cm$^{-2}$. The ionization yield is smaller than in the case of the resonant wavelength at the same laser pulse parameters (Fig. 2(b,d)), while they are nearly the same at $7.5\times10^{13}$ W cm$^{-2}$ (see upper panels in Fig. 2). Since the states that have been ionized are the same, the reason may be only in their population which overwhelms a small difference of ionization rates in favor of a longer wavelength ($\sigma_n^i(E_L) \propto [(IP - E_n)/E_L]^{7/2}$ according to the Kramers formula).

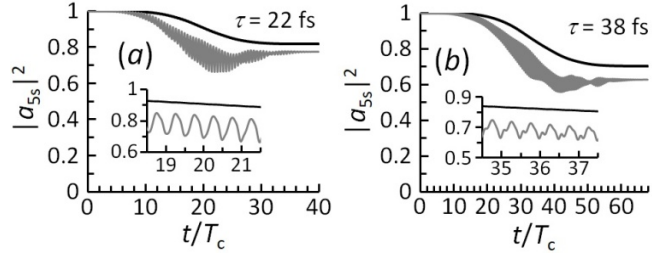

Fig. 7. The same as in Fig. 2, but at the laser wavelength 460 nm and the peak intensity $I_p = 1.5\times10^{14}$ W cm$^{-2}$. (a) Pulse duration $\tau = 22$ fs; (b) $\tau = 38$ fs.

Very short (namely, sub-femtosecond, or attosecond) XUV pulses can be obtained from a group of HH [2-4]. Attosecond pulse generation using a broad group of resonantly enhanced harmonics in Xe was found theoretically in [34]. Pronounced resonant enhancement of a single harmonic is not compatible with the attosecond pulse generation. However, moderate enhancement of the resonant HH found in our calculations under moderate laser intensities makes attosecond pulse production feasible.

We calculate the XUV field as follows:

$$E_{XUV}(t) = \int_\Omega^\infty E_\omega \exp(-i\,\omega t) d\omega, \qquad (8)$$

where $\Omega$ is the lower limit of the used frequency range, and the XUV field spectral amplitude is defined by $E_\omega \propto \omega \mathcal{F}_\omega[\mathbf{D}_{av}(t)] \cdot \mathbf{e}$.

The XUV intensity $|E_{XUV}(t)|^2$ found via Eq. (8) using $\Omega = 4\omega_L$ is shown in Fig. 8. One can see that the train of the attosecond pulses evolves at the femtosecond scale. The insets show most intense fragments of the train.

The attosecond pulse generation shows the phase-locking of the harmonics, including the resonant one. The FWHM of the attosecond pulse generated in the resonant case is 190 as. The inverse bandwidth of the XUV spectrum is about 150 as (the spectrum for this case is shown in Fig. 4(c)). So the attosecond pulse duration found in our calculations is close to the spectrally-limited one, showing almost perfect synchronization of HH generated in the moderate laser field near the laser pulse maximum.

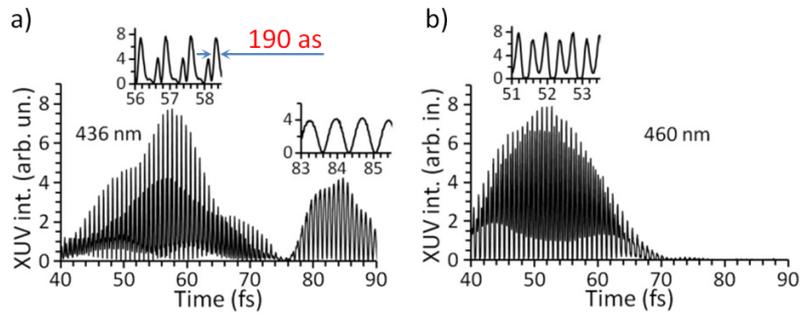

Fig. 8. The attosecond pulse trains calculated via Eq. (8) for $\tau = 38$ fs and $I_p = 7.5 \times 10^{13}$ W cm$^{-2}$, laser wavelength 436 nm, i. e. exact resonance with the AIS (a) and laser wavelength 460 nm (b). The insets show attosecond pulses near maxima of the trains.

## 4. Conclusions

We calculate the HHG spectra for the In$^+$ ion in the field of intense femtosecond laser pulse. The model is based on expansion of the TDSE solution over a finite set of bound eigenstates of ionic Hamiltonian. The time-dependent expansion coefficients are used to obtain the average dipole moment for the emission spectra in the framework of semiclassical approach.

Our calculations show the enhancement of high harmonic intensity under the multiphoton resonance for transition from the ground state to the AIS $4d^95p$ $^1P_1$. Moreover, we demonstrate the attosecond pulse train generation using high harmonics including the resonant one. The attosecond pulse duration found for the moderate laser intensity is close to the XUV field inverse bandwidth and is as short as 190 as.

Our findings of the resonant HHG enhancement are supported by the recent similar result for the Ga$^+$ ion [33] and those obtained previously in frame of the multiphoton perturbation theory [32]. Note that under high laser intensities our calculations show lower enhancement magnitude than the one observed experimentally [9].

We expect the reason in more accurate account of continuum states, which will be the subject of the consideration in the near future.

## Funding

This study was funded by the RSF (Grant No. 22-12-00389).

## Disclosures

The authors declare no conflict of interest.